\documentclass[]{IEEEtran}

\usepackage{amsmath,amsfonts}
\usepackage{algorithmic}
\usepackage{algorithm}
\usepackage{array}
\usepackage[caption=false,font=normalsize,labelfont=sf,textfont=sf]{subfig}
\usepackage{textcomp}
\usepackage{stfloats}
\usepackage{url}
\usepackage{verbatim}
\usepackage{graphicx}
\usepackage[pdftex,pdfpagelabels,bookmarks,hyperindex,hyperfigures]{hyperref}
\usepackage{cite}
\usepackage{color}
\usepackage{colortbl}
\usepackage{tabularx}
\usepackage{multirow}
\usepackage{cleveref}
\usepackage{booktabs}
\usepackage{tabularx}
\usepackage{pifont}

\definecolor{lightblue}{rgb}{0.78,0.85,0.95}

\begin{document}
	\title{Network Digital Twin for Open RAN: The Key Enablers, Standardization, and Use Cases}
	
	\author{Javad Mirzaei, Ibrahim Abualhaol, Gwenael Poitau\\
		\small {Advanced Wireless Technology, Dell Technologies Inc.}
		\thanks{This work is developed within the Advanced Wireless Technology (AWT), AI/ML at Dell Technologies Inc. \texttt{Emails}: \texttt{Javad.Mirzaei@Dell.com, Ibrahim.Abualhaol@Dell.com, Gwenael.Poitau@Dell.com}.}
	}

	\maketitle
	
	\begin{abstract}
		 \label{sec:Abs}
The open radio access network (O-RAN), with its disaggregated and open architecture, is poised to meet the demands of the next generation of wireless communication. However, to unlock the full potentials of O-RAN, real-time network modeling and optimization are essential. 
A promising solution for such requirement is the use of network digital twin (NDT). NDT provides a comprehensive view of a network, covering both physical and logical components, including infrastructure, protocols, and algorithms. NDT, as a real-time virtual representation of O-RAN facilitates a variety of operations, such as emulations, test, optimization, monitoring, and analysis of a new configuration in a risk-free environment, without requiring them to be implemented in real network. Such capability enables the vendors and network operators for a faster adoption of new solutions with frequent updates, while ensuring the resiliency of the existing services via planning ahead under various “what-if” scenarios.  In this paper, we first describe what exactly NDT means in the context of O-RAN, as well as its key enablers. We then describe the NDT application within the O-RAN in both prior and post-deployment. Finally, we provide two practical uses cases, namely network energy efficiency and traffic steering, where the NDT can be leveraged effectively.

	\end{abstract}
	
\begin{IEEEkeywords}
	5G,
	Real-time modeling and optimization,
	Network Digital Twin,
	O-RAN
\end{IEEEkeywords}
\section{Introduction}  \label{sec:Intro}

The telecommunications industry is currently undergoing substantial changes, fueled by the rapid advances in wireless networks. These enhancements aim to support the emergence and expectations of new wireless technologies, notably 5G and those yet to come. This fast-paced transformation necessitates innovative solutions like the open radio access network (O-RAN) architecture, an avant-garde approach catering to a wide array of requirements, from ultra-low latency communication (URLLC) and massive machine-type communication (mMTC), to enhanced mobile broadband (eMBB). Historically, telecommunications networks have been predominantly singular-vendor, proprietary systems. O-RAN disrupts this norm by proposing an open architecture that nurtures flexibility, cost-effectiveness, and swift innovation. This shift is facilitated through the separation of hardware and software components, artificial intelligence (AI), virtualization, disaggregation, as well as promoting a multi-vendor ecosystem that bolsters dynamic services and competition.

Despite the O-RAN's potential to revolutionize the telecommunications industry, it's deployment has several challenges. Traditionally, the testing and assurance of RAN functions conducted in a lab setting in an offline manner before being implemented in real network followed by frequent updates. This approach is quite time-consuming, and requires a careful planning to manage the process. In the context of O-RAN, such an offline network management is not practical, mainly due to the network heterogenity, and the challenges related to the multi-vendor interoperability driven by disaggregated nature of the O-RAN. In particular, the offline techniques are incapable of delivering a comprehensive, standardized, and end-to-end testing framework, which is the key to ensure the O-RAN's reliability and guaranteed performance. To unleash the full potentials of O-RAN, the industry demands for a resilient, and yet scalable solution in order to automatically test, validate, and optimize the O-RAN in real-time both in prior and post-deployment.

In light of the above challenges, this paper aims to provide a detailed examination of the role of network digital twins (NDT) in addressing the challenges in O-RAN. We first provide a concrete discussion on what exactly NDT means in O-RAN and its key enablers. Our contribution here is mainly on how to leverage the NDT in O-RAN's deployment in both prior and post-deployment. We then provide a high level view on the integration of NDT in various O-RAN use cases. As an example, we delve into two practical O-RAN use cases - traffic steering and energy efficiency - offering sequence diagrams and data types, to show its operational efficiency.

The work in this paper strives to shed light on the importance and the potential of NDTs in transforming network management within the O-RAN context. By addressing the associated challenges and elucidating the benefits and practical considerations of implementing NDTs into future-generation networks, we aim to pave the way for their successful application and realization. Ultimately, our research aspires to significantly contribute to the transformation and optimization of network management processes, guaranteeing the smooth and efficient functioning of O-RAN.

The remaining of this paper is organized as follows: Section \ref{sec:O-RANTechnologyReview} provides an overview of the O-RAN technology including its key features, components and interfaces. In Section \ref{sec:NDT_for_ORAN}, we first describe what NDT means in the context of O-RAN, where we provide a highlight on the efforts made within the standard, as well as a comparison between the NDT and its alternatives. We then explore on how to integrate the NDT into the O-RAN operations followed by the key enablers to realize such an integration. In Section \ref{sec:Practical_NDT}, we describe the practical use cases of NDT in O-RAN from the vendors and operators' perspective, mainly in prior and post-deployment of O-RAN. Section \ref{sec:conclusion}, provides the concluding remarks.

\section{O-RAN Technology Overview} \label{sec:O-RANTechnologyReview}

The existing RAN is composed of a set of monolithic components that implement the protocols in each layer of cellular communication in an end-to-end manner. Importantly, these components are provided by only a limited number of vendors. Relying on such an architecture, limits the usage of the existing RAN to meet the demands in 5G and beyond. In particular, the limited reconfigurability and inter-coordination among the network components of the RAN prevents the joint optimization and control of the RAN components, leading to a poor support for diverse traffic profiles driven by various use cases. Under such circumstances, the network operators have to continuously maintain and upgrade their network infrastructure to keep up with the market trends and to meet the demands to be made by their customers\cite{O-RAN_IEEE}. Therefore, adopting the existing RAN for the next generation of cellular communication is costly both from the Capex and Opex perspectives.

To address the above challenges, the O-RAN Alliance, a consortium of academic and industry members, has pushed a new paradigm for the future RAN. According to \cite{O-RAN.WG1.O-RAN-Architecture-Description-v07.00}, The O-RAN is architecturally designed based on the following key principles in mind: open interfaces, virtualization, and intelligence. The benefit of such open architecture is in multi-folds. For example, the open interfaces principle emphasizes the use of standardized interfaces between the network elements to enable interoperability and reduced vendor lock-in. It further  enables a flexible RAN deployment, which allows even small players to operate within this ecosystem. The virtualization principle aims to maximize resource utilization and flexibility by decoupling the RAN functions from the underlying hardware stack. The intelligence principle is considered to enhance the performance and efficiency of the network through the power of machine learning (ML) algorithms.  
	
\begin{figure}[t]
	\centering
	\includegraphics[width=1\linewidth]{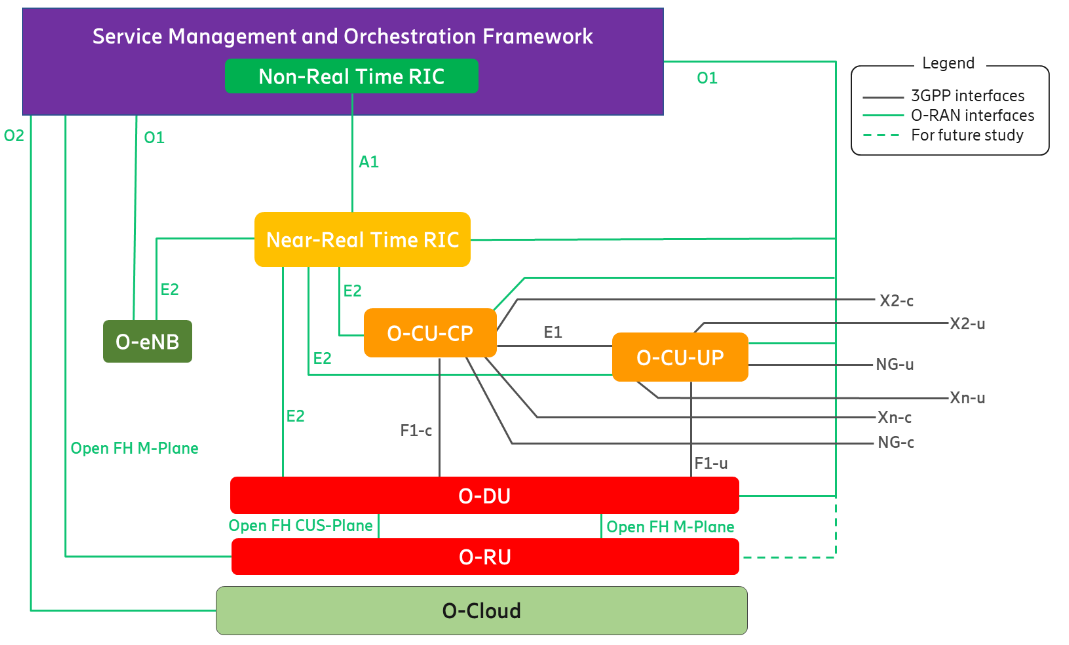}
	\caption{Logical Architecture of O-RAN \cite{O-RAN.WG1.O-RAN-Architecture-Description-v07.00}}
	\label{fig:oran-logical-architecture}
\end{figure}

	As shown in Fig. \ref{fig:oran-logical-architecture},  the logical architecture is comprised of five key components: the radio unit (RU), the distributed unit (DU), the central unit (CU), and the non-real time radio intelligence controller (RIC) and near-real time RIC. The RU is responsible for the radio interface and lower physical layer processing of the radio signal, while the DU is responsible for the upper physical layer processing and medium access control (MAC), and acts as a bridge between the RU and the CU. The CU component is mainly in charge of the core network functions, including the user plane and control plane. The non-real time RIC and near-real time RIC are dedicated to optimize network performance and automate network management. In particular, the non-real time RIC is responsible for long-term optimization of the RAN, while the near-real time RIC is responsible for faster (near real-time) optimization. The service management and orchestration (SMO) framework  provides centralized management and orchestration of the RAN, and it enables the intelligence principle by using the ML techniques to automate service management and orchestration. Note that, these components are aligned with the O-RAN principles of open interfaces, virtualization, and intelligence. These components are communicating via the standardized and open interfaces, which give rise to interoperability between different vendors' hardware and software, reducing vendor lock-in as well as increasing collaboration among different entities.
	
	\section{NDT for 5G O-RAN} \label{sec:NDT_for_ORAN}
	
	\subsection{What does NDT mean for O-RAN in 5G?}

	The surge in 5G adoption has imposed a considerable demand on operators and vendors to streamline the construction, integration, and management of these networks. The O-RAN, with its AI-enabled, disaggregated and open architecture, is poised to meet the needs of the next generation of cellular communication such as the URLLC, mMTC, and eMBB. The full potentials of such heterogeneous applications in O-RAN can only be unlocked through a real-time optimization, test, and validation.
	
	Traditionally, the testing and assurance of services and/or protocols are conducted in a physical lab setting in an offline manner before  being implemented in real network followed by occasional updates. Such an approach is costly and time-consuming, requiring a careful planning to manage the process. In the context of O-RAN, the offline network management is infeasible, mainly due to i) disaggregated nature of the O-RAN, incurring challenges in multi-vendor interoperability, software/model updates, leading to performance degradation, and ii) the network heterogeneity and its associated complexities, including bandwidth, latency and massive connectivity. Therefore, testing and assurance in O-RAN demands for a real-time approach, which can be done via an emulation on a replica of an intended RAN in real-time. Such a real-time process allows for an efficient and faster deployment, as well as an accurate testing and monitoring of the next-generation network.

NDT, as a true digital replica of O-RAN,  provides a sandbox to try various "what-if" scenarios, identify risks and deficiencies, and test various risk mitigation strategies, that enables proactive network management for the optimal network performance. Such capability offers a rapid prototyping, testing, and validation, thereby reducing time-to-market while ensuring  seamless integration and interoperability of O-RAN components.
 
 As O-RAN continues to evolve and grow in size and complexity, the NDT, as a real-time virtual representation of O-RAN, facilitates a variety of network use cases, such as emulations, analysis and diagnosis in a safe and zero-risk environment, without requiring them to be implemented in real network. While such capabilities play an increasingly important role in ensuring the efficiency, reliability, and sustainability of O-RAN, the NDT is envisioned as a mean to achieve the zero-touch networking.
 
NDT can also be leveraged to generate large amounts of high-quality data that can be used to train AI/ML algorithms. Such data will be used to develop predictive models that can help automate network management and optimize network performance.  

\subsection{The Standard Perspective for NDT}
From the standard perspective, there is limited effort on establishing a standardized NDT for O-RAN. The existing efforts are mainly focused on O-RAN test specifications. For example, 3GPP provides a set of 5G specifications for testing that covers functional, performance, and conformance tests based on the specific 3GPP protocols. ETSI, on a work item under the generic automatic network architecture (GANA) program, proposes a general framework for testing and validating the AI-integrated networks, including data, algorithm, and model validation, as well as non-functional and integration testing \cite{ETSI_WP}. While the focus of these efforts is mainly on AI-enabled systems/networks, they do not provide specifics about the network requirements, management tasks, or AI models.  ETSI emphasizes on the need for AI system testing and defines a generic process, however it is not aligned with the O-RAN architecture. In particular, it does not specify how to test and verify the O-RAN specific interfaces, the xApps/rApp in RIC, and the network functions. O-RAN Alliance work group (WG) 4, in a closely-related effort, extends the 3GPP standards and defines test cases, parameters, and procedures for testing the conformance and performance of the O-DU, O-CU, and O-RU \cite{o-ran-E2E_tests}. The O-RAN Alliance Test and Integration focus group (FG), specifies the scope, goals, and processes for an end-to-end network testing, where the O-RAN system under the test is treated as a black box. Such process is suffering from a certain deficiencies, which demands for an NDT-based O-RAN architecture that enables a real-time optimization, test, and validation.

\begin{table*}[t]
	\centering
	\renewcommand*\arraystretch{1.1}
	\caption{Comparison of Simulators, Emulators, and NDTs in O-RAN.}
	\label{tab:comparison_NDT}
	\resizebox{\textwidth}{!}{%
		\begin{tabular}{p{2.5cm}|p{4.2cm}|p{4.5cm}|p{4.3cm}}
			\hline
			\textbf{Attribute} & \textbf{Simulator} & \textbf{Emulator} & \textbf{NDT} \\
			\hline
			Scope & Network-level Test \& Validation & Component-level Test \& Optimize & Comprehensive Network Mgmt. \\
			\hline
			Environment & Virtual & Virtual and Physical & Virtual \\
			\hline
			Real-Time & None & None & Enabled by design\\
			\hline
			End-to-End Testing & Yes & Limited& Enabled by design\\
			\hline
			Inputs & Network Parameters & Network \& Component Parameters & Component-level KPIs \& Configs \\
			\hline
			Outputs & Network-level KPIs & Component-level KPIs & Comprehensive KPIs \\
			\hline
			Scalability & Limited & Limited & Enabled by design\\
			\hline
			Generalizability & Limited & Limited & Enabled by design\\
			\hline
			{Speed} & {Non Real-time}  &  {Real-time with limitations}  & {Real-time} \\
			\hline
		\end{tabular}}
\end{table*}
\subsection{NDT vs Simulator and Emulator for O-RAN}
In order to optimize and test the O-RAN, one can use an emulator to mimic the behavior of specific components, such as O-CU, O-DU or O-RU. However, the emulation technique is costly to implement especially for a sizable O-RAN, and poses limitations for an online test and validation. Alternatively, simulators can create a virtualized environment to model the behaviors of a component or a process. However, the simulators do not fully represent the real-world O-RAN mainly due to modeling deficiencies. Importantly, employing the simulators are both timely and computationally expensive to be used for O-RAN use cases. In contrast, an NDT, leveraging the state-of-the-art ML techniques, provides a more comprehensive model of the entire network in real-time, making it easier to be used in O-RAN applications.

It is worth mentioning that the NDT and its physical counterpart (i.e., the O-RAN) require to have a real-time  communication. This is the main difference between the NDT and its close alternatives such as simulator and emulator. Unlike the simulator and emulator, where they have no interaction over the real physical environments, the NDT connects with the physical assets and tries to represent it with little-to-no assumption and/or simplification \cite{9220177}. Table \ref{tab:comparison_NDT} provides a comprehensive comparision between the NDT and its alternatives.

\subsection{The Integration of NDT in O-RAN }
At the high level, as shown in Fig. \ref{fig:NDT_Comp}, NDT-enabled O-RAN architecture is based on the following three pillars: i) The Physical Twin (PT), which is the RAN physical network including the RU, DU, and CU. The PT is the source of data,  ii) The NDT, which is aimed to represent the replica of its physical counterpart in  real-time. The NDT is the host of  various components such as the basic and functional models (which will be described later in this section), the data repository and the NDT management sub-components, and iii) The real-time data flow between the PT and its digital counterpart (i.e., the southbound (SB)), and the northbound (NB) communication between the NDT and RIC. To achieve the full potentials of NDT in O-RAN, there are several key enablers, that we itemize as follows:

\begin{figure*}[t]
	\centering
	\includegraphics[width=1.0\linewidth]{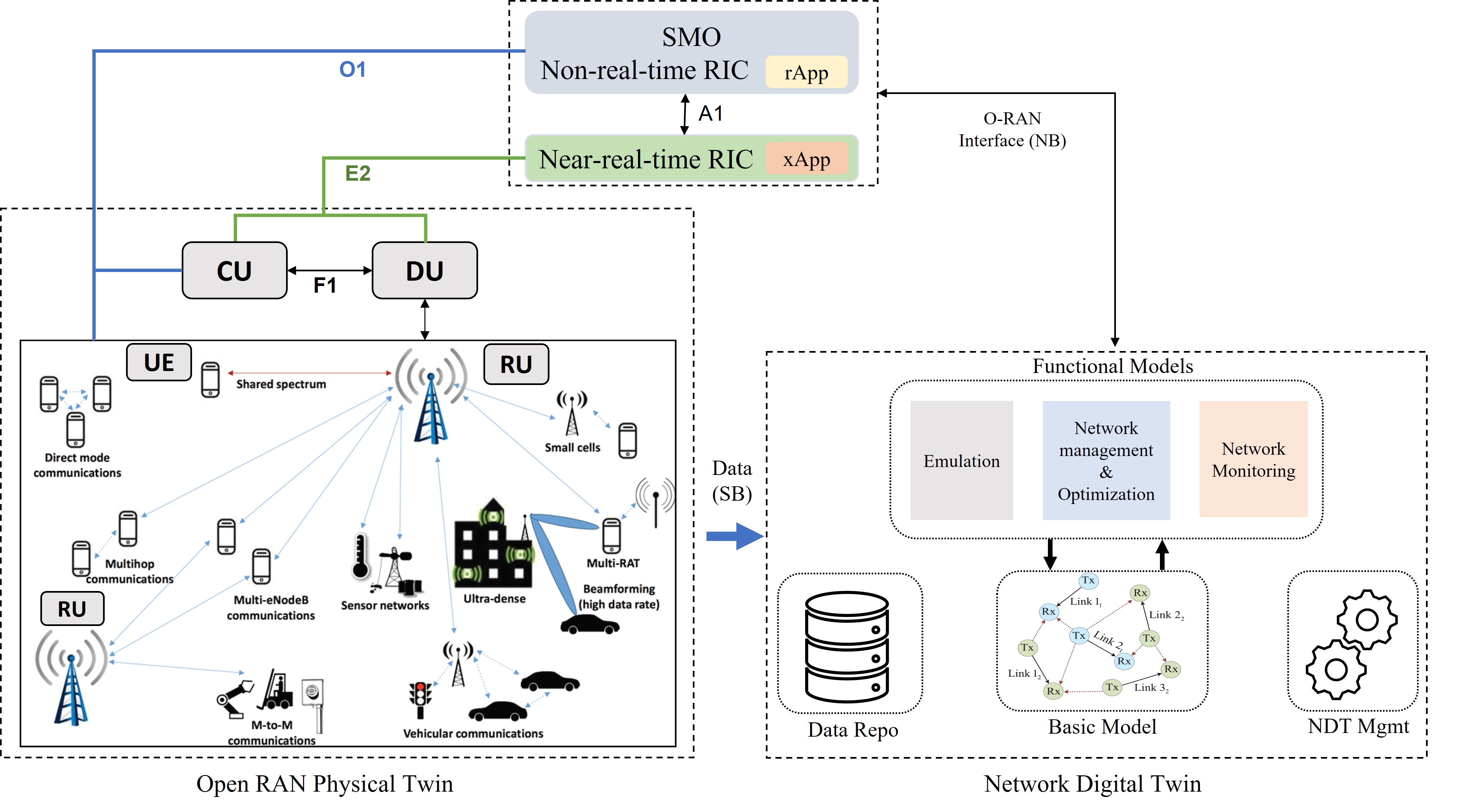}
	\caption{The NDT integration in 5G O-RAN.}
	\label{fig:NDT_Comp}
\end{figure*}

\begin{itemize}
	\item \emph{Standardized Interfaces and Protocols}: The standard interfaces and protocols ensures that different components of the network can interoperate and communicate with each other, even if they are from different vendors. Such capability enables network operators to choose among the best solutions for their network operations, without relying on a single vendor. Additionally, as the network scales up, the integration of new components with the existing network becomes straightforward, without requiring significant changes to the network architecture. Besides, due to the heterogeneity of services and devices envisioned in O-RAN, the modular construction of NDT is inevitable. This not only allows for an efficient scaling of the NDT, but also, facilitates the multi-vendor NDT development. However, the key enabler of such modular design pattern remains in standard and vendor-agnostic interfaces between the modules.
	\item \emph{Data Strategy}: Developing a high-fidelity NDT requires a comprehensive set of data collected from the physical network, while maintaining a balance between the quality and the quantity of the data. This demands a careful assessment on the type of data collection to avoid the unnecessary redundancies, and save on communication overhead and storage resources. The data collection mechanism has to be aligned with the existing O-RAN standardized interfaces such as E2, that encompasses different service models under the E2SM \cite{O-RAN.WG3.E2SM-RC-R003-v03.00} and E2AP \cite{O-RAN.WG3.E2AP-R003-v03.01} protocols.  Fig. \ref{fig:NDT_E2} provides an illustration on how NDT can leverage the E2 interfaces to collect the required data from the RAN.

It is also critical to recognize the significance of a reliable data management to maintain the long-term stability and adaptability of NDT for O-RAN. Such a framework should cover a complete life-cycle of data, starting from its collection, storage, maintenance, and retrieval. To achieve an optimal data management, automation is highly desirable, along with other essential features such as security, accurate record-keeping, traceability, and data integrity. These features ensure the proper handling of sensitive network data, while maintaining the accuracy, completeness and consistency of data.

	\item \emph{Modeling}: The models are used as a proxy to represent different entities of the PT, which enable various tasks, including analyzing, diagnosing, and emulation in its digital counterpart.  The models are generally categorized as basic models and functional models. The basic models are mainly used to replicate the O-RAN components, such as the UEs, gNBs, eNBs, channels, topology and network slices, while the functional models are used to enable various network-level functionalities such as emulations, prediction, network management, and monitoring. It is worth noting that the fidelity of NDT is highly related to the accuracy of such models to capture the network complexities. That is, the ML techniques can be leveraged in developing such models.
	\begin{figure}[ht]
		\centering
		\includegraphics[width=.95\linewidth]{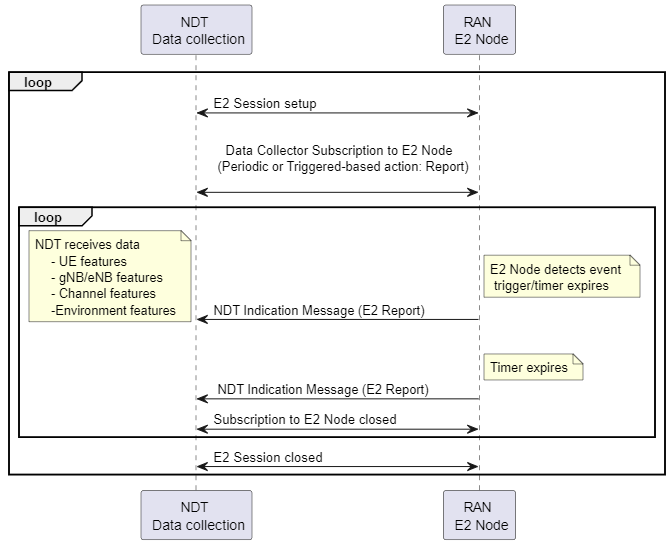}
		\caption{Call flow between the NDT data collectors and E2 Node in O-RAN. }
		\label{fig:NDT_E2}
	\end{figure} 
	\item \emph{Deployment}: Depending on the intended use cases in O-RAN, NDT can be deployed either at the edge, cloud, or their combination. The choice of the deployment is a balance between the  delay requirement and the available computation power. For example, the edge-based deployment is characterized by lower computational power and lower storage capability, and it is mainly used in delay-sensitive URLLC applications, such as real-time monitoring, performance optimization, and fault detection. On the other hand, the cloud-based deployment is  mainly used for network-wide planning, slicing, optimization, and troubleshooting.
\end{itemize}

	\section{Practical use cases of NDT in O-RAN}\label{sec:Practical_NDT}

	From the vendors and operators' perspectives,
	development and operation of O-RAN is a significant challenge mainly due to its virtualized, disaggregated, and multi-vendor design. To address such challenges, NDT can be leveraged in both prior and post-deployment of O-RAN. In the former stage, NDT allows for extensive testing and validation, while in the later stage, the NDT facilitates the real-time optimization and operation across different domains (access, transport, and core). In the following section, we will elaborate on these use cases in further details.
	
	\subsection{NDT for Prior-deployment of O-RAN}
	At this stage, the NDT provides a true replica of the network environment to plan, test, and validate the design of new applications and services in O-RAN \cite{9696282}. In particular, the NDT can assess the operations of different components, applications and/or services against a variety of “what-if” scenarios (e.g., traffic loads) that have never been experienced before. Doing so, the vendors ensure the integrity and robustness of their intended services to meet the demands of a hypothetical situation before they are deployed on a live network. This provides a risk-free process to seamlessly onboard different services in O-RAN. 
	
	The integration of ML techniques in O-RAN is extensively covered in the literature \cite{O-RAN_IEEE}, \cite{ORAN_Alliance1}. The authors in \cite{O-RAN_IEEE} provide a detailed introduction of different components in O-RAN, and then highlights how these components will enable the integration of ML techniques as a closed-loop control mechanisms in RIC. However, the successful adoption of the ML techniques is tied to an accurate training of ML models prior to deployment. This is where the NDT can play a critical role in training of such models. For example, in the case of reinforcement learning (RL), the algorithm is trained via a trial-and-error mechanism by constantly interacting with an environment. Indeed, the algorithm needs to explore various choices in the action space to learn the best policy that optimizes the objective function in a long run. Such a continuous trial-and-error process in not affordable in real physical network. The NDT provides a safe and realistic-like environment to train such models and enable faster integration of RL techniques in production. The authors of \cite{s23031197} proposed a detailed framework to leverage the NDT to train an RL algorithm for capacity sharing use case.

	\subsection{NDT for Post-deployment of O-RAN}
	
	In a live O-RAN, multiple vendors generate agile individual network functions that require to push updates frequently across the RU, DU, and CUs. Traditionally, every vendor has to go through a testing process on a simulator or emulator which is costly to setup and provides no guarantee on the accuracy of the outcome. Importantly, in order to ensure the relevancy of the tests in a multi-vendor setting, each vendor has to track the compatibility of their service with others, which is not feasible. With the O-RAN roll-out, it then becomes necessary to ensure an automatic and continuous operational assessment that validates interoperability, standard conformance among other performance certificates for any changes made by the vendors. This requires  that O-RAN to integrate into the vendors' continuous-integration continuous-development (CI/CD) pipeline.  The NDT, as a true and real-time digital replica of the live O-RAN, provides an environment to automatically test and validate the integrity and performance of the function updates. It also facilitates to proactively identify the potential performance degradation and their corresponding root causes before the function update is pushed to the live network. For example, NDTs can be leveraged in network monitoring and real-time anomaly detection when the network deviates from its normal operations or to predict any service disruptions before they even happen. The integration of ML techniques in this context will further enhance the  mean-time-to-response (MTTR). In Table \ref{table:NDT_for_Prior_Post}, we summarize the application of NDT for prior and post-deployment of O-RAN.

	\begin{table*}[htbp]
		\centering
		\renewcommand*\arraystretch{1.3}
		\caption{NDT for Prior and Post-deployment of O-RAN. Use Cases are from \cite{oran_UseCase_Report_10}. }
		\label{table:NDT_for_Prior_Post}
		\resizebox{\textwidth}{!}{%
		\begin{tabular}{p{1.8cm}|p{6.3cm}| p{1.5cm} p{1.5cm} p{1.5cm} p{1.3cm} }
			\hline
			\hfil{\textbf{Deployment}} & \hfil{\multirow{2}{*}{\textbf{Use Cases}}} &  \multicolumn{4}{c}{\textbf{NDT Application}}   \\
			\cline{3-6}
			\hfil{\textbf{Stage}}&  &\hfil{Planing} & \hfil{Operation} & \hfil{AI/ML} & \hfil{Monitoring} \\
			\hline
			& O-RAN-based Industrial IoT &  \hfil \checkmark &  &   &   \\
			& Local Indoor Positioning in RAN  &  \hfil \checkmark &   & \hfil \checkmark &   \\
			& Application/Service design &  \hfil \checkmark &   & \hfil \checkmark  &   \\
			\cline{2-2}
				& Massive MIMO Beamforming Optimization & \hfil \checkmark &  \hfil \checkmark &  \hfil \checkmark &  \\
				& MIMO SU/MU-MIMO Optimization & \hfil \checkmark &  \hfil \checkmark &  \hfil \checkmark &   \\
	\hfil Prior & QoE/QoS Based Resource Optimization& \hfil \checkmark &  \hfil \checkmark &  \hfil \checkmark &   \\
				& Energy Efficiency & \hfil \checkmark &  \hfil \checkmark &  \hfil \checkmark &  \hfil \checkmark \\
				& Network Slicing  &  &  \hfil \checkmark &  \hfil \checkmark &   \hfil \checkmark\\
				& Traffic Steering &  &  \hfil \checkmark & \hfil \checkmark  &  \hfil \checkmark \\
				& Dynamic Spectrum Sharing (DSS) &   &  \hfil \checkmark & \hfil \checkmark  & \hfil \checkmark  \\
				& Dynamic UAV Radio Resource Allocation &   &  \hfil \checkmark &  \hfil \checkmark & \hfil \checkmark  \\
				\cline{1-1}
				& BBU Pooling for RAN Elasticity  &  &  \hfil \checkmark &  \hfil \checkmark &  \hfil \checkmark \\
				& Cross-Domain Orchestration & \hfil \checkmark &  \hfil \checkmark & \hfil \checkmark  &  \hfil \checkmark \\
				& Context-Based Dynamic Handover &  &  \hfil \checkmark & \hfil \checkmark  & \hfil \checkmark  \\
				& Dynamic RAN Sharing &  &  \hfil \checkmark &   \hfil \checkmark&  \hfil \checkmark \\
				& Shared O-RU &  &  \hfil \checkmark &  \hfil \checkmark & \hfil \checkmark  \\
				& Application/Services Validation &  \hfil \checkmark &  \hfil \checkmark & \hfil \checkmark  &  \hfil \checkmark \\
				& xApp/rApp Model Training, Testing \& Validation  & \hfil \checkmark &   &  \hfil \checkmark &   \\
				& Policy Design & \hfil \checkmark &  &  \hfil \checkmark &   \\
	\hfil Post	& Calibration &  & \hfil \checkmark &  \hfil \checkmark &   \\
			\cline{2-2}
			& RAN Slice SLA Assurance &   &  \hfil \checkmark &  \hfil \checkmark &  \hfil \checkmark \\
			& Security Threads Assessment  &   & \hfil \checkmark & \hfil \checkmark  &  \hfil \checkmark\\
			& KPIs report: Resources Utilization, Throughput &   & \hfil \checkmark &  \hfil \checkmark & \hfil \checkmark \\
			& Process \& Visualize the Network Performance  &   & \hfil \checkmark & &  \hfil \checkmark \\
			& Congestion Prediction \& Management &   & \hfil \checkmark & \hfil \checkmark  &  \hfil \checkmark \\
			& Anomaly Detection  &   & \hfil \checkmark &  \hfil \checkmark &  \hfil \checkmark \\
			\hline
		\end{tabular}}
	\end{table*}

	It is worth noting that, there are various O-RAN testing and validation platforms developed by either universities or industry labs with different levels of maturity. However, not all these  frameworks fully support the testing and planning of the O-RAN in both prior and post-deployment. In particular, the existing solutions are based on emulations and/or simulations which limit their capabilities for a comprehensive, automated, and (near-) real-time test and validation of a sizable O-RAN. In addition, the existing solutions can only serve a partially deployed O-RAN, which may be sufficient for the purpose of research and development. However, vendors, operators, and other ecosystem players who are involved in the business of building or operating O-RAN as a service need to focus far more deeply on component, system-level testing, as well as the cross-vendor interoperability assessments in order to realize an efficient O-RAN deployment. Importantly, NDT's holistic view effectively addresses use cases across multiple domains, such as access, transport, core, and application, in O-RAN. Its multi-domain capability enables seamless cross-domain orchestration, enhancing network resource utilization and performance.
	
\subsection{Use Cases}
\begin{figure*}[hbt!]
	\centering
	\includegraphics[width=.77\linewidth]{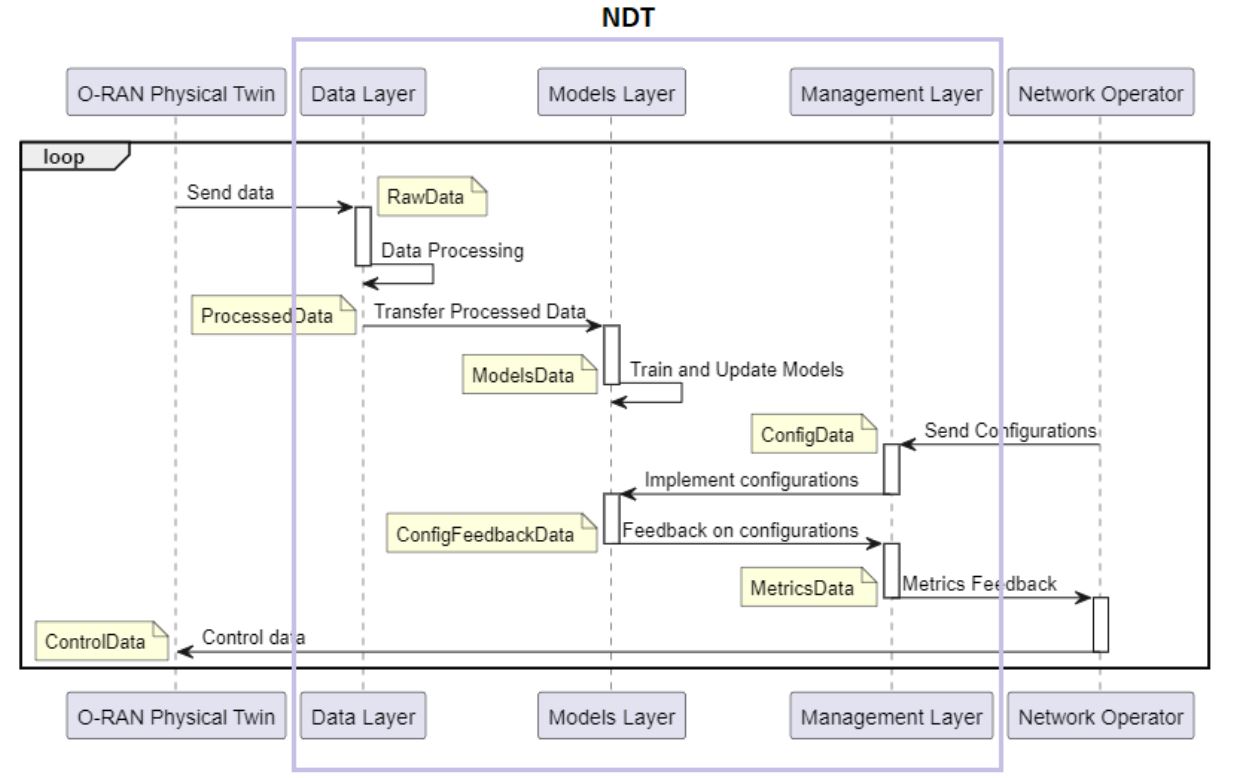}
	\caption{Data Flow between the O-RAN physcal twin,  NDT, and network operator for various use cases in O-RAN.}
	\label{fig:usecaseseq}
\end{figure*}

In this section, we provide two important use cases in O-RAN where the NDT can be leveraged. For each use case, the common process between the NDT and its physical counterpart (i.e., the O-RAN) is depicted in Fig. \ref{fig:usecaseseq}. This process involves the following key components: the O-RAN Physical Twin, the Data Layer, the Models Layer, the Management Layer, and the Network Operator. The detailed description of each data type for the sample use cases are provided in Table  \ref{tab:usecases_data_types}.

\subsubsection{NDT for Traffic Steering}
Traffic Steering is a use case that involves dynamically directing traffic in the network to optimize performance and manage network loads. Within the O-RAN, the proposed NDT architecture can be leveraged to collect the real-time data from the network and create a digital model that can simulate different traffic scenarios. The NDT will help network operators to make informed decisions about optimizing traffic flow and managing network loads in order to satisfy the required QoS/QoE. Additionally, the NDT can be utilized to test various network configurations and identify potential issues before they happen in the live network. 
\subsubsection{ NDT for Energy Efficiency}
In the context of O-RAN, energy efficiency aims to optimize energy usage within the network. This can be accomplished by implementing techniques such as deactivating unused equipment during periods of low traffic, optimizing power utilization in active devices, or adjusting the flow of network traffic. The main challenge is to find the right compromise between energy and QoS KPIs. Relying on its real-time capabilities, NDT can be leveraged to test and validate various network configurations and assess their efficacy from the energy consumption perspective, before they get pushed to the live network.
\begin{table}[h]
	\centering
	\renewcommand*\arraystretch{1.05}
	\caption{Use case specific data type description.}
	\label{tab:usecases_data_types}
	\resizebox{\columnwidth}{!}{%
		\begin{tabular}{p{1.8cm}|p{3.08cm}|p{2.7cm}}
			\hline
			\multirow{2}{*}{\textbf{Data Type}} &  \multicolumn{2}{c}{\textbf{Use Case}}   \\
			\cline{2-3}
			& \textbf{Energy Efficiency} & \textbf{Traffic Steering}\\
			\hline
			\multirow{2}{*}{{RawData}} &  Energy Consumption & Traffic Load \\
			& Traffic Load/Type & User Mobility Pattern \\
			\hline
			\multirow{2}{*}{{ProcessedData}} &  Energy Usage &  Traffic Demand\\
			&  Traffic Demand & User Mobility\\
			\hline
			{{ModelsData}} &  Energy Consumption  & Traffic Demands  \\
			(Predictions)&  Traffic Abnormality  & Routing Decisions \\
			\hline
			\multirow{2}{*}{{ConfigData}} &  Energy Saving Config. & Traffic Steering Rules \\
			&  Power Allocation & Load-Balancing Param. \\
			\hline
			{{ConfigFeedback}} &  \multicolumn{2}{c}{\multirow{2}{*}{{ Compliance Configuration}}}  \\
			\hfil Date&\multicolumn{2}{c}{ }\\
			\hline
			\multirow{2}{*}{{MetricsData}} & Energy Usage & Throughput, Delay \\
			& Energy Saving & Traffic Distribution \\
			\hline
			\multirow{3}{*}{{ControlData}} &  Power Control & Traffic Re-Routing \\
			& Carrier/Cell Switch On/Off & Load Balancing \\
			& Sleep Modes Config. & Handover Trigger \\
			\hline
	\end{tabular}}
\end{table}

	\section{Conclusion} \label{sec:conclusion}
	NDTs have the potential to transform the telecommunications industry by providing real-time modeling, and optimization of the next-generation of wireless networks. As a digital replica of the O-RAN, NDT enables vendors and network operators to emulate, test, and optimize their intended services under various "what-if" scenarios in a risk-free environment, without requiring them to be implemented in real network. 
	
	Throughout this paper, we provided an overview on how NDT can be leveraged in the context of O-RAN. We described the architecture of NDT within the O-RAN operation, and the key enablers of such integration. We also provided a comprehensive discussion on the practical application of NDT in various O-RAN use cases, including the both prior and post-deployment. Furthermore, network energy efficiency and traffic steering are provided as example uses cases in O-RAN where we provided a detailed description on how NDT can be integrated effectively to satisfy the intended QoS/QoE.

	\bibliographystyle{IEEEtran}  
	\bibliography{references}

	\vfill

\end{document}